\documentclass{article}
\usepackage{times}
\usepackage{amsfonts}
\usepackage{graphicx}
\usepackage[pdfmark]{hyperref}
\begin{document}
\noindent
{\Large    QUANTUM MECHANICS AS A SPONTANEOUSLY\\ BROKEN  GAUGE THEORY ON A U(1) GERBE}
\vskip1cm
\noindent
{\bf Jos\'e M. Isidro}\\
Grupo de Modelizaci\'on Interdisciplinar, Instituto de Matem\'atica Pura y Aplicada,\\ Universidad Polit\'ecnica de Valencia, Valencia 46022, Spain\\
Max--Planck--Institut f\"ur Gravitationsphysik, Albert--Einstein--Institut,\\ D--14476 Golm, Germany\\
{\tt joissan@mat.upv.es}
\vskip1cm
\noindent
\today
\vskip1cm
\noindent
{\bf Abstract} Any quantum--mechanical system possesses a U(1) gerbe naturally defined on configuration space. Acting on Feynman's kernel $\exp({\rm i}S/\hbar)$, this U(1) symmetry allows one to arbitrarily pick the origin for the classical action $S$, on a point--by--point basis on configuration space. This is equivalent to the statement that quantum mechanics is a U(1) gauge theory. Unlike Yang--Mills theories, however, the geometry of this gauge symmetry is not given by a fibre bundle, but rather by a gerbe.
Since this gauge symmetry is spontaneously broken, an analogue of the Higgs mechanism must be present. We prove that a Heisenberg--like noncommutativity for the space coordinates is responsible for the breaking. This allows to interpret the noncommutativity of space coordinates as a Higgs mechanism on the quantum--mechanical U(1) gerbe.

\tableofcontents

\section{Introduction}\label{bernabeudemierdeu}

Let $\mathbb{M}$ be an $n$--dimensional spacetime manifold endowed with the the metric tensor $g_{\mu\nu}$. Let $x^{\mu}$, $\mu=1,\ldots, n$, be local coordinates on $\mathbb{M}$. The possibility of measuring the infinitesimal distance 
\begin{equation}
{\rm d}s^2=g_{\mu\nu}{\rm d}x^{\mu}{\rm d}x^{\nu}
\label{kkaa}
\end{equation}
between two points on $\mathbb{M}$ rests on the assumption that the corresponding coordinates can be simultaneously measured with infinite accuracy, so one can have
\begin{equation}
\Delta x^{\mu}=0
\label{rda}
\end{equation}
simultaneously for all $\mu=1,\ldots, n$. In quantum--mechanical language one would recast this assumption as 
\begin{equation}
[\hat x^{\mu},\hat x^{\nu}]=\hat x^{\mu}\hat x^{\nu}-\hat x^{\nu}\hat x^{\mu}=0,
\label{ccss}
\end{equation}
where $\hat x^{\mu}$ is a quantum operator whose classical limit is the coordinate function $x^{\mu}$. The vanishing of the above commutator expresses two alternative, though essentially equivalent, statements, one of physical content, the other geometrical. Physically it expresses the absence of magnetic fields across the $\mu,\nu$ directions \cite{GMS}. Geometrically it expresses the fact that the multiplication law on the algebra of functions on the space $\mathbb{M}$ is commutative.

All modern theories of quantum gravity \cite{QUANTUMGRAVITY} share the common feature that a minimal length scale, the {\it Planck length}\/ $L_P$, exists on spacetime, 
\begin{equation}
\Delta x^{\mu}\geq L_P,
\label{bbcc}
\end{equation}
so $L_P$ effectively becomes the shortest possible distance, and its square $L_P^2$ becomes proportional to the quantum of area.
This coarse graining of a spacetime continuum $\mathbb{M}$ can be mimicked, in noncommutative geometry \cite{NCG}, by noncommuting operator coordinates $\hat x^{\mu}$ acting as Hermitean operators on Hilbert space $\mathbb{H}$. The $\hat x^{\mu}$ satisfy
\begin{equation}
[\hat x^{\mu},\hat x^{\nu}]={\rm i}a\theta^{\mu\nu},
\label{cghp}
\end{equation}
with $\theta^{\mu\nu}$ a constant, real, dimensionless antisymmetric tensor. Here $a>0$ is a fundamental area scale, such that
\begin{equation}
\lim_{a\to 0}[\hat x^{\mu},\hat x^{\nu}]=0.
\label{rktdpk}
\end{equation}
Moreover, in the limit $a\to 0$, one can identify (possibly up to some singular renormalisation factor $Z$) the operator $\hat x^{\mu}$ on $\mathbb{H}$ with the function $x^{\mu}$ on $\mathbb{M}$. Since the Heisenberg uncertainty relations corresponding to (\ref{cghp}) imply
\begin{equation}
\Delta\hat x^{\mu}\Delta\hat x^{\nu}\geq \frac{a}{2}\vert\theta^{\mu\nu}\vert, 
\label{rkbr}
\end{equation}
the above statement concerning the coarse graining of $\mathbb{M}$ follows. Up to possible numerical factors $C$ one can therefore set
\begin{equation}
a=CL_P^2.
\label{rzf}
\end{equation}

It has been argued \cite{PADUNO} that the existence of a fundamental length scale $L_P$ on $\mathbb{M}$ implies modifying the spacetime metric according to the rule
\begin{equation}
{\rm d}s^2\longrightarrow{\rm d}s^2+L_P^2,
\label{rkcak}
\end{equation}
so $L_P$ effectively becomes the shortest possible distance. One can also prove \cite{PADUNO} that modifying the spacetime interval according to (\ref{rkcak}) is equivalent to requiring invariance of a field theory under the following exchange of short and long distances:
\begin{equation}
{\rm d}s\longleftrightarrow\frac{L_P^2}{{\rm d}s}.
\label{paddy}
\end{equation}
{}Further consequences of the exchange (\ref{paddy}) have been reported in ref. \cite{PADOS}.

On the other hand, we have in ref. \cite{ME} shown that the existence of a minimal length scale $L_P$ is equivalent to the exchange
\begin{equation}
\frac{S}{\hbar}\longleftrightarrow\frac{\hbar}{S}
\label{rchp}
\end{equation}
in Feynman's exponential of the action integral $S$:
\begin{equation}
\exp\left({\rm i}\,\frac{S}{\hbar}\right)\longleftrightarrow\exp\left({\rm i}\,\frac{\hbar}{S}\right).
\label{bbmk}
\end{equation}
In other words, the duality (\ref{paddy}) is equivalent to the duality (\ref{bbmk}). Since the equations of motion that follow from the variation of $S/\hbar$ are the same as those derived from the variation of $\hbar/S$, classically there is no difference between $S/\hbar$ and $\hbar/S$.  We will refer to the exchange (\ref{rchp}) as {\it semiclassical vs. strong--quantum duality}. This simple $\mathbb{Z}_2$--transformation has been extended \cite{ME} to larger duality groups $G$ such as ${\rm SL}\,(2,\mathbb{Z})$, ${\rm SL}\,(2,\mathbb{R})$  and ${\rm SL}\,(2,\mathbb{C})$. Examples of the semiclassical {\it vs.}\/ strong--quantum duality (\ref{rchp}) have appeared under different, though essentially equivalent, guises, in refs. \cite{PINZULSTERNGAONA}; see \cite{VARIOUS} for related works. 

Examining the relation between the noncommutativity (\ref{cghp}) of the space coordinates and the quantum of area (\ref{rzf}) one realises that eqns. (\ref{cghp}) and (\ref{rzf}) are in fact equivalent.  The commutation relations (\ref{cghp}) imply the existence of a quantum of area: by (\ref{cghp}) one has $\Delta\hat x^{j}\sim L_P$, hence a quantum of area must exist and be proportional to $L_P^2$. Conversely, let a quantum of area $\Delta\hat x^{j}\Delta\hat x^{k}\sim L_P^2$ be given. The latter could not exist on a spacetime continuum whose coordinates all commute, since then we would always have $\Delta\hat x^{j}\Delta\hat x^{k}=0$.  The simplest noncommutativity giving rise to a quantum of area is (\ref{cghp}); more general types of noncommutativity can also be considered \cite{LIZZI}. The above equivalence between  eqns. (\ref{cghp}) and (\ref{rzf}) is intuitively obvious, but it will be very instructive to recast it in the geometrical language of gerbes \cite{ORLANDO}. We have in ref. \cite{NOI} succeeded in interpreting quantum mechanics as a U(1) gauge theory on phase space. However, unlike Yang--Mills theories, the gauge symmetry is not expressed geometrically by means of a connection 1--form and its corresponding curvature 2--form on a fibre bundle. Rather, the appropriate geometrical setup will be provided by a gerbe. 

The first goal of this paper is {\it to break the  U(1) symmetry on the quantum--mechanical gerbe}\/ constructed in ref. \cite{NOI}. The breaking will occur via an analogue of the usual Higgs mechanism of Yang--Mills theory, as adapted now to the fact that  gerbes live one step up from bundles (for the geometrical aspects of the Higgs mechanism see ref. \cite{SARDANASHVILY}). This breaking is necessary since the exchange (\ref{paddy}), or its equivalent (\ref{bbmk}) on the U(1) gerbe, is certainly not realised in Nature as observed at low energies. However dualities such as (\ref{paddy}) and (\ref{bbmk}) are to be expected \cite{PADUNO} within the realm of quantum gravity. Moreover, quantum--gravity effects have also been conjectured to be relevant at astrophysical scales \cite{REUTER}; gravity itself can be understood as arising from the breaking of local Lorentz symmetry \cite{POTTING}. All this evidence strongly suggests a study of the symmetry--breaking mechanism in our setup.

As a second goal of this article, we will prove that {\it  a space noncommutativity of the type (\ref{cghp}) provides the gerbe analogue of the Higgs mechanism in Yang--Mills theory}. Our previous results of ref. \cite{NOI} were deduced on phase space, where an interesting link could be established with the phase--space formulation of quantum mechanics \cite{GOSSON}. In the present paper we will work on configuration space instead.

To summarise, we will see that the requirement of semiclassical {\it vs.} strong--quantum duality (\ref{bbmk}) will lead to a quantisation of spacetime, and viceversa. Thus the gerbe approach to nonrelativistic quantum mechanics analysed here, duly generalised to the relativistic case, can provide an interesting route towards a quantum theory of gravity.

\section{A gerbe on configuration space}\label{chpbbln}

In ref. \cite{NOI} we have given a detailed construction of a quantum--mechanical gerbe on phase space. In what follows we briefly recall its main features and adapt it to configuration space.

\subsection{Basics in gerbes}\label{rman}

It is well known \cite{BOTTTU} that a unitary line bundle on a base manifold $\mathbb{M}$ is a 1--cocycle $\lambda\in H^1\left(\mathbb{M}, C^{\infty}({\rm U}(1))\right)$. The latter is the first \v Cech cohomology group of $\mathbb{M}$ with coefficients in the sheaf of germs of smooth, U(1)--valued functions. Let $\left\{U_{\alpha}\right\}$ be a good cover of $\mathbb{M}$ by open sets $U_{\alpha}$. Then the bundle is determined by a collection of U(1)--valued transition functions defined on each 2--fold overlap
\begin{equation}
\lambda_{\alpha_1\alpha_2}:U_{\alpha_1}\cap U_{\alpha_2}\longrightarrow {\rm U}(1)
\label{marsalis}
\end{equation}
satisfying 
\begin{equation}
\lambda_{\alpha_2\alpha_1}=\lambda^{-1}_{\alpha_1\alpha_2}, 
\label{winton}
\end{equation}
as well as the 1--cocycle condition
\begin{equation}
\lambda_{\alpha_1\alpha_2}\lambda_{\alpha_2\alpha_3}\lambda_{\alpha_3\alpha_1}=1\quad {\rm on}\quad U_{\alpha_1}\cap U_{\alpha_2}\cap U_{\alpha_3}.
\label{ptbb}
\end{equation}

A {\it gerbe}\/  is defined as a 2--cocycle $g\in H^2\left(\mathbb{M}, C^{\infty}({\rm U}(1))\right)$. This means that we have a collection $\{g_{\alpha_1\alpha_2\alpha_3}\}$ of maps defined on each 3--fold overlap on $\mathbb{M}$
\begin{equation}
g_{\alpha_1\alpha_2\alpha_3}:U_{\alpha_1}\cap U_{\alpha_2}\cap U_{\alpha_3}\longrightarrow {\rm U}(1)
\label{bbmkmd}
\end{equation}
satisfying
\begin{equation}
g_{\alpha_1\alpha_2\alpha_3}=g^{-1}_{\alpha_2\alpha_1\alpha_3}=g^{-1}_{\alpha_1\alpha_3\alpha_2}=g^{-1}_{\alpha_3\alpha_2\alpha_1},
\label{ktdpkbb}
\end{equation}
as well as the 2--cocycle condition
\begin{equation}
g_{\alpha_2\alpha_3\alpha_4}\,g^{-1}_{\alpha_1\alpha_3\alpha_4}\,g_{\alpha_1\alpha_2\alpha_4}\,g^{-1}_{\alpha_1\alpha_2\alpha_3}=1\quad {\rm on}\quad U_{\alpha_1}\cap U_{\alpha_2}\cap U_{\alpha_3}\cap U_{\alpha_4}.
\label{knptbb}
\end{equation}
Now $g$ is a 2--coboundary in \v Cech cohomology whenever it holds that
\begin{equation}
g_{\alpha_1\alpha_2\alpha_3}=\tau_{\alpha_1\alpha_2}\tau_{\alpha_2\alpha_3}\tau_{\alpha_3\alpha_1}
\label{lag}
\end{equation}
for a certain collection $\left\{\tau_{\alpha_1\alpha_2}\right\}$ of U(1)--valued functions $\tau_{\alpha_1\alpha_2}$ on $U_{\alpha_1}\cap U_{\alpha_2}$ such that $\tau_{\alpha_2\alpha_1}=\tau^{-1}_{\alpha_1\alpha_2}$. The collection $\left\{\tau_{\alpha_1\alpha_2}\right\}$ is called a {\it trivialisation}\/ of the gerbe. One can prove that over any given open set $U_{\alpha}$ of the cover $\left\{U_{\alpha}\right\}$ there always exists a trivialisation of the gerbe. 

On a gerbe specified by the 2--cocycle $g_{\alpha_1\alpha_2\alpha_3}$, a connection is specified by forms $A, B, H$ satisfying
\begin{eqnarray}
H\vert_{U_{\alpha}}&=&{\rm d}B_{\alpha}\\
B_{\alpha_2}-B_{\alpha_1}&=&{\rm d}A_{\alpha_1\alpha_2}\\
A_{\alpha_1\alpha_2}+A_{\alpha_2\alpha_3}+A_{\alpha_3\alpha_1}&=&g^{-1}_{\alpha_1\alpha_2\alpha_3}{\rm d}g_{\alpha_1\alpha_2\alpha_3}.
\label{ktpyy}
\end{eqnarray}
The 3--form $H$ is the curvature of the gerbe connection. The latter is called {\it flat}\/ if $H=0$.

\subsection{The trivialisation}\label{vtr}

Let an action integral $S$ be given for a point particle on the spacetime $\mathbb{M}$. Let us further assume that the latter factorises, at least locally, as the product of the time axis $\mathbb{R}$ and a configuration space $\mathbb{F}$. Coordinates $x_{(\alpha)}^{\mu}$ on the local chart labelled by ${\alpha}$ therefore decompose as $(t_{\alpha}, q_{\alpha}^{j})$, with $j=1$, $\ldots$, $n-1$. This latter index will be suppressed in what follows. Let any two points $q_{\alpha_1}$, $q_{\alpha_2}$ be given on $\mathbb{F}$, with local charts $U_{\alpha_1}$, $U_{\alpha_2}$ centred around them. Charts for the time coordinate will not be indicated explicitly unless necessary. Moreover, let $\mathbb{L}_{\alpha_1\alpha_2}$ be an oriented path connecting $q_{\alpha_1}$ to $q_{\alpha_2}$ as time runs from $t_{\alpha_1}$ to $t_{\alpha_2}$. We define $a_{\alpha_1\alpha_2}$ as the following functional integral over all such trajectories $\mathbb{L}_{\alpha_1\alpha_2}$:
\begin{equation}
a_{\alpha_1\alpha_2}\sim\int {\rm D}\mathbb{L}_{\alpha_1\alpha_2}\exp\left[\frac{{\rm i}}{\hbar}S(\mathbb{L}_{\alpha_1\alpha_2})\right].
\label{bbkkp}
\end{equation}
Throughout this paper, the $\sim$ sign  will stand for {\it proportionality}: path integrals are defined up to some (usually divergent) normalisation. However all such normalisation factors will cancel in the ratios of path integrals that we are interested in. The argument of the exponential in eqn. (\ref{bbkkp}) contains the action $S$ evaluated along the path $\mathbb{L}_{\alpha_1\alpha_2}$. Thus $a_{\alpha_1\alpha_2}$ is proportional to the probability amplitude for the particle to start at $q_{\alpha_1}$ and finish at $q_{\alpha_2}$, {\it i.e.}, it is proportional to the propagator $G(q_{\alpha_1}, t_{\alpha_1};q_{\alpha_2}, t_{\alpha_2})$:
\begin{equation}
\int {\rm D}\mathbb{L}_{\alpha_1\alpha_2}\exp\left[\frac{{\rm i}}{\hbar}S(\mathbb{L}_{\alpha_1\alpha_2})\right]\sim G(q_{\alpha_1}, t_{\alpha_1};q_{\alpha_2}, t_{\alpha_2}). 
\label{puteubernabeu}
\end{equation}
Now assume that $U_{\alpha_1}\cap U_{\alpha_2}$ is nonempty,
\begin{equation} 
U_{\alpha_1\alpha_2}:=U_{\alpha_1}\cap U_{\alpha_2}\neq \phi.
\label{neptyy}
\end{equation}
and define, for $(q_{\alpha_{12}}, q_{\alpha_1}, q_{\alpha_2})\in U_{\alpha_1\alpha_2}\times U_{\alpha_1}\times U_{\alpha_2}$,
$$
\tau'_{\alpha_1\alpha_2}\colon  U_{\alpha_1\alpha_2}\times U_{\alpha_1}\times U_{\alpha_2} \longrightarrow \mathbb{C}
$$
\begin{equation}
\tau'_{\alpha_1\alpha_2}:=a_{\alpha_1\alpha_{12}}a_{\alpha_{12}\alpha_2}.
\label{csgkdh}
\end{equation}
Thus $\tau'_{\alpha_1\alpha_2}$ is proportional to the probability amplitude for the following transition: starting at $q_{\alpha_1}$, the particle reaches $q_{\alpha_2}$ {\it after}\/ traversing the variable midpoint $q_{\alpha_{12}}$. We have
\begin{equation}
\tau'_{\alpha_1\alpha_2}\sim\int{\rm D}\mathbb{L}_{\alpha_1\alpha_2}(\alpha_{12})\exp\left\{\frac{{\rm i}}{\hbar}\,S\left[\mathbb{L}_{\alpha_1\alpha_2}(\alpha_{12})\right]\right\}
\label{beth}
\end{equation}
$$
=\int{\rm D}\mathbb{L}_{\alpha_1\alpha_{12}}\exp\left[\frac{{\rm i}}{\hbar}S(\mathbb{L}_{\alpha_1\alpha_{12}})
\right]\int{\rm D}\mathbb{L}_{\alpha_{12}\alpha_{2}}\exp\left[\frac{{\rm i}}{\hbar}S(\mathbb{L}_{\alpha_{21}\alpha_{2}})\right]
$$
$$
\sim G(q_{\alpha_1}, t_{\alpha_1};q_{\alpha_{12}}, t_{\alpha_{12}}) G(q_{\alpha_{12}}, t_{\alpha_{12}};q_{\alpha_2}, t_{\alpha_2}).
$$
As it stands,  $\tau'_{\alpha_1\alpha_2}$ is a function on $U_{\alpha_1\alpha_2}\times U_{\alpha_1}\times U_{\alpha_2}$ because of its dependence on the endpoints $q_{\alpha_1}$ and $q_{\alpha_2}$, which are being kept fixed. A true trivialisation should be a function on the double overlap $U_{\alpha_1\alpha_2}$ only. However we can integrate $\tau'_{\alpha_1\alpha_2}$ over $q_{\alpha_1}$ and $q_{\alpha_2}$ in order to eliminate this dependence. We thus define
\begin{equation}
\tilde\tau_{\alpha_1\alpha_2}\colon  U_{\alpha_1\alpha_2} \longrightarrow\mathbb{C}, 
\label{bona}
\end{equation}
$$
\tilde\tau_{\alpha_1\alpha_2}:=\int{\rm d}q_{\alpha_1}{\rm d}q_{\alpha_2}\tau'_{\alpha_1\alpha_2}
$$
$$
=\int{\rm d}q_{\alpha_1}{\rm d}q_{\alpha_2}G(q_{\alpha_1}, t_{\alpha_1};q_{\alpha_{12}}, t_{\alpha_{12}}) G(q_{\alpha_{12}}, t_{\alpha_{12}};q_{\alpha_2}, t_{\alpha_2}).
$$
Since a trivialisation must be a U(1)--valued function, we finally define
\begin{equation}
\tau_{\alpha_1\alpha_2}\colon  U_{\alpha_1\alpha_2} \longrightarrow{\rm U(1)}, \qquad
\tau_{\alpha_1\alpha_2}:=\frac{\tilde\tau_{\alpha_1\alpha_2}}{\vert \tilde\tau_{\alpha_1\alpha_2}\vert},
\label{bernabeumekagueuentuputamadreu}
\end{equation}
whenever $\tilde\tau_{\alpha_1\alpha_2}$ is nonvanishing. One can verify that  $\tau_{\alpha_1\alpha_2}$ qualifies as a trivialisation on $\mathbb{F}$. Physically, this trivialisation is interpreted as the U(1)--valued phase of the probability amplitude for the particle to start at {\it any}\/ initial point in the chart $U_{\alpha_1}$ and to reach {\it any}\/ final point in the chart $U_{\alpha_2}$, while traversing the midpoint $q_{\alpha_{12}}\in U_{\alpha_1\alpha_2}$. Observe that (\ref{bona}) contains a Riemann volume integral while (\ref{beth}) contains a Feynman path integral. Notice also that  (\ref{bernabeumekagueuentuputamadreu}) depends parametrically on the times $t_{\alpha_1}$, $t_{\alpha_2}$ and $t_{\alpha_{12}}$; it will also depend parametrically on whatever other parameters the action $S$ may contain such as masses, forces, frequencies, coupling constants, {\it etc}. However, as the trivialisation of a gerbe over $\mathbb{F}$, $\tau_{\alpha_1\alpha_2}$ depends only on the point $q_{\alpha_{12}}\in U_{\alpha_1\alpha_2}$ as it should. 

The trivialisations corresponding to a number of cases are worked out explicilty in the appendix. These examples prove that, at least up to (and including) quadratic terms, which is the degree of approximation we will keep throughout, whatever zeroes the propagators may have, these zeroes will all cancel in the end. Thus the trivialisation, being a U(1)--phase,  is always well defined. One can think of $\tau_{\alpha_1\alpha_2}(q_{\alpha_{12}})$ as the U(1)--phase of the averaged ({\it i.e.}, integrated) probability amplitude for the particle to start somewhere in $U_{\alpha_1}$ and finish somewhere in $U_{\alpha_2}$ while crossing $q_{\alpha_{12}}\in U_{\alpha_1\alpha_2}$.

\subsection{The 2--cocycle}\label{dos}

Next consider three points and their respective charts
\begin{equation}
q_{\alpha_1}\in U_{\alpha_1},\qquad q_{\alpha_2}\in U_{\alpha_2}, \qquad q_{\alpha_3}\in U_{\alpha_3}
\label{ttppo}
\end{equation}
such that the triple overlap $U_{\alpha_1}\cap U_{\alpha_2}\cap U_{\alpha_3}$ is nonempty,
\begin{equation}
U_{\alpha_1\alpha_2\alpha_3}:=U_{\alpha_1}\cap U_{\alpha_2}\cap U_{\alpha_3}\neq \phi.
\label{pacca}
\end{equation}
Once the trivialisation (\ref{bona}) is known, the 2--cocycle $g_{\alpha_1\alpha_2\alpha_3}$ defining a gerbe on $\mathbb{F}$ is given by  (\ref{lag}):
$$
g_{\alpha_1\alpha_2\alpha_3}\colon U_{\alpha_1\alpha_2\alpha_3}\longrightarrow {\rm U(1)}
$$
\begin{equation}
g_{\alpha_1\alpha_2\alpha_3}(q_{\alpha_{123}}):=\tau_{\alpha_1\alpha_2}(q_{\alpha_{123}})\tau_{\alpha_2\alpha_3}(q_{\alpha_{123}})\tau_{\alpha_3\alpha_1}(q_{\alpha_{123}}),
\label{eptcg}
\end{equation}
where all three $\tau$'s on the right--hand side are, by definition, evaluated at the same variable midpoint  
\begin{equation}
q_{\alpha_{123}}\in U_{\alpha_1\alpha_2\alpha_3}.
\label{vef}
\end{equation}
Being U(1)--valued, the 2--cocycle (\ref{eptcg}) can be expressed as the quotient of a complex function $\tilde g$ by its modulus,
\begin{equation}
g_{\alpha_1\alpha_2\alpha_3}(q_{\alpha_{123}})=\frac{\tilde g_{\alpha_1\alpha_2\alpha_3}(q_{\alpha_{123}})}{\vert \tilde g_{\alpha_1\alpha_2\alpha_3}(q_{\alpha_{123}})\vert}.
\label{miqq}
\end{equation}
By eqns. (\ref{bona}) and (\ref{eptcg}) we have
\begin{equation}
\tilde g_{\alpha_1\alpha_2\alpha_3}(q_{\alpha_{123}})
\label{bernabeumafioseu}
\end{equation}
$$
\sim\int{\rm d}q_{\alpha_1}{\rm d}q_{\alpha_2}G(q_{\alpha_1},t_{\alpha_1};q_{\alpha_{123}},t_{\alpha_{123}})G(q_{\alpha_{123}},t_{\alpha_{123}};q_{\alpha_2},t_{\alpha_2})
$$
$$
\times \int{\rm d}q'_{\alpha_2}{\rm d}q_{\alpha_3}G(q'_{\alpha_2},t'_{\alpha_2};q_{\alpha_{123}},t'_{\alpha_{123}})G(q_{\alpha_{123}},t'_{\alpha_{123}};q_{\alpha_3},t_{\alpha_3})
$$
$$
\times \int{\rm d}q'_{\alpha_3}{\rm d}q'_{\alpha_1}G(q'_{\alpha_3},t'_{\alpha_3};q_{\alpha_{123}},t''_{\alpha_{123}})G(q_{\alpha_{123}},t''_{\alpha_{123}};q'_{\alpha_1},t'_{\alpha_1}),
$$
where
\begin{equation}
t_{\alpha_1}<t_{\alpha_{123}}<t_{\alpha_2}<t'_{\alpha_2}<t'_{\alpha_{123}}<t_{\alpha_3}<t'_{\alpha_3}<t''_{\alpha_{123}}<t'_{\alpha_1}.
\label{tempus}
\end{equation}
Thus $g_{\alpha_1\alpha_2\alpha_3}(q_{\alpha_{123}})$ equals the U(1)--phase of the probability amplitude for the following transition: starting anywhere in $U_{\alpha_1}$ (say, at $q_{\alpha_1}$), the particle crosses  $q_{\alpha_{123}}$ on its way to some $q_{\alpha_2}\in U_{\alpha_2}$; the points $q_{\alpha_1}$ and $q_{\alpha_2}$ are integrated over. Next, starting at some $q'_{\alpha_2}\in U_{\alpha_2}$, the particle crosses the same $q_{\alpha_{123}}$ again on its way to some $q_{\alpha_3}\in U_{\alpha_3}$; the points $q'_{\alpha_2}$ and $q_{\alpha_3}$ are also integrated over. Finally, from $q'_{\alpha_3}\in U_{\alpha_3}$ it traverses $q_{\alpha_{123}}$ once more before finally reaching some $q'_{\alpha_1}\in U_{\alpha_1}$; the  points $q'_{\alpha_3}$ and $q'_{\alpha_1}$ are also integrated over. 

It must be observed that the points $q'_{\alpha_1}$, $q'_{\alpha_2}$ and $q'_{\alpha_3}$ are not necessarily identical with $q_{\alpha_1}$, $q_{\alpha_2}$ and $q_{\alpha_3}$, respectively. Thus the transition considered does not necessarily define a closed path on $\mathbb{F}$, although all such paths traverse $q_{\alpha_{123}}$.
Moreover, condition (\ref{tempus}) implies that the complete trajectory is never closed as a path on $\mathbb{F}\times\mathbb{R}$. However, in the particular case that one or more of the equalities $q_{\alpha_1}=q'_{\alpha_1}$,  $q_{\alpha_2}=q'_{\alpha_2}$ and $q_{\alpha_3}=q'_{\alpha_3}$ does {\it not}\/ hold, we can always connect the points $q_{\alpha_j}$ and $q'_{\alpha_j}$ within the corresponding $U_{\alpha_j}$, so as to complete a closed loop on $\mathbb{F}$. This closed loop is the projection, onto $\mathbb{F}$, of an open loop on $\mathbb{F}\times\mathbb{R}$. It is possible to complete such an open path to a closed loop because the transition amplitides considered above are all integrated over the endpoints $q_{\alpha_j}$ and $q'_{\alpha_j}$. In so doing we obtain a closed loop on $\mathbb{F}$ such as that in the figure:
\begin{equation}
\mathbb{L}_{\alpha_1\alpha_2\alpha_3}(\alpha_{123}):=\mathbb{L}_{\alpha_1\alpha_2}(\alpha_{123})+\mathbb{L}_{\alpha_2\alpha_3}(\alpha_{123})+\mathbb{L}_{\alpha_3\alpha_1}(\alpha_{123}).
\label{fuga}
\end{equation}
Recalling that the propagator can be expressed as the functional integral (\ref{puteubernabeu}), we conclude that (\ref{bernabeumafioseu}) can be expressed as a functional integral over all closed loops on $\mathbb{F}$ of the type (\ref{fuga}):
\begin{equation}
\tilde g_{\alpha_1\alpha_2\alpha_3}(q_{\alpha_{123}})\sim\int{\rm D}\mathbb{L}_{\alpha_1\alpha_2\alpha_3}(\alpha_{123})\exp\left\{\frac{{\rm i}}{\hbar}\,S\left[\mathbb{L}_{\alpha_1\alpha_2\alpha_3}(\alpha_{123})\right]\right\}.
\label{xam}
\end{equation}
{}From now on we will restrict our attention to closed loops on $\mathbb{F}$ of the type (\ref{fuga}).\footnote{The above discussion also settles an apparent discrepancy between the definition of the trivialisation given here and that given in ref. \cite{NOI}. The correct definition of the trivialisation is the one given in section \ref{vtr} here. However the 2--cocycle  (\ref{miqq}), (\ref{xam}) obtained from the trivialisation of section \ref{vtr},  and therefore the gerbe itself, coincides with that of ref. \cite{NOI}.}

Next we will recast eqn. (\ref{xam}) into an equivalent, but more useful, expression. Given a closed loop $\mathbb{L}$, let $\mathbb{S}\subset\mathbb{F}$ be a 2--dimensional surface with boundary such that $\partial\mathbb{S}=\mathbb{L}$. By Stokes' theorem,
\begin{equation}
S(\mathbb{L})=\int_{\mathbb{L}}{\cal L}{\rm d}t=\int_{\partial\mathbb{S}}{\cal L}{\rm d}t=\int_{\mathbb{S}}{\rm d}{\cal L}\wedge{\rm d}t.
\label{toes}
\end{equation}
Any surface $\mathbb{S}$ such that $\partial\mathbb{S}=\mathbb{L}$ will satisfy eqn. (\ref{toes}) because the integrand ${\rm d}{\cal L}\wedge{\rm d}t$ is closed. Let us now choose $\mathbb{S}$ to bound a closed loop $\mathbb{L}_{\alpha_1\alpha_2\alpha_3}(\alpha_{123})$ as in eqn. (\ref{fuga}). Consider the first half of the leg $\mathbb{L}_{\alpha_1\alpha_2}(\alpha_{123})$, denoted $\frac{1}{2}\mathbb{L}_{\alpha_1\alpha_2}(\alpha_{123})$. The latter runs from $\alpha_1$ to $\alpha_{123}$. Consider also the second half of the leg $\mathbb{L}_{\alpha_3\alpha_1}(\alpha_{123})$, denoted $\frac{1}{2'}\mathbb{L}_{\alpha_3\alpha_1}(\alpha_{123})$, with a prime to remind us that it is the {\it second}\/ half: it runs back from $\alpha_{123}$ to $\alpha_1$. The sum of these two half legs,
\begin{equation}
\frac{1}{2}\mathbb{L}_{\alpha_1\alpha_2}(\alpha_{123})+\frac{1}{2'}\mathbb{L}_{\alpha_3\alpha_1}(\alpha_{123}),
\label{mezzo}
\end{equation}
completes one roundtrip and it will, as a rule, enclose an area $\mathbb{S}_{\alpha_1}(\alpha_{123})$, unless the path from $\alpha_{123}$ to $\alpha_1$  happens to coincide exactly with the path from $\alpha_1$ to $\alpha_{123}$:
\begin{equation}
\partial\mathbb{S}_{\alpha_1}(\alpha_{123})=\frac{1}{2}\mathbb{L}_{\alpha_1\alpha_2}(\alpha_{123})+\frac{1}{2'}\mathbb{L}_{\alpha_3\alpha_1}(\alpha_{123}).
\label{soprano}
\end{equation}
Analogous conclusions apply to the other half legs $\frac{1}{2'}\mathbb{L}_{\alpha_1\alpha_2}(\alpha_{123})$, $\frac{1}{2}\mathbb{L}_{\alpha_3\alpha_1}(\alpha_{123})$, $\frac{1}{2}\mathbb{L}_{\alpha_2\alpha_3}(\alpha_{123})$ and $\frac{1}{2'}\mathbb{L}_{\alpha_2\alpha_3}(\alpha_{123})$ under cyclic permutations of 1,2,3 in the \v Cech indices $\alpha_1$, $\alpha_2$ and $\alpha_3$:
\begin{equation}
\partial\mathbb{S}_{\alpha_2}(\alpha_{123})=\frac{1}{2}\mathbb{L}_{\alpha_2\alpha_3}(\alpha_{123})+\frac{1}{2'}\mathbb{L}_{\alpha_1\alpha_2}(\alpha_{123}),
\label{tenor}
\end{equation}
\begin{equation}
\partial\mathbb{S}_{\alpha_3}(\alpha_{123})=\frac{1}{2}\mathbb{L}_{\alpha_3\alpha_1}(\alpha_{123})+\frac{1}{2'}\mathbb{L}_{\alpha_2\alpha_3}(\alpha_{123}).
\label{bajo}
\end{equation}
The boundaries of the three surfaces $\mathbb{S}_{\alpha_1}(\alpha_{123})$, $\mathbb{S}_{\alpha_2}(\alpha_{123})$ and $\mathbb{S}_{\alpha_3}(\alpha_{123})$ all pass through the variable midpoint $\alpha_{123}$, although we will no longer indicate this explicitly. We define their connected sum
\begin{equation}
\mathbb{S}_{\alpha_1\alpha_2\alpha_3}:=\mathbb{S}_{\alpha_1}+\mathbb{S}_{\alpha_2}+\mathbb{S}_{\alpha_3}.
\label{wdvdd}
\end{equation}
In this way we have
\begin{equation}
\mathbb{L}_{\alpha_1\alpha_2\alpha_3}=\partial\mathbb{S}_{\alpha_1\alpha_2\alpha_3}=
\partial\mathbb{S}_{\alpha_1}+\partial\mathbb{S}_{\alpha_2}+\partial\mathbb{S}_{\alpha_3}.
\label{bbvvdd}
\end{equation}
It must be borne in mind that $\mathbb{L}_{\alpha_1\alpha_2\alpha_3}$ is a function of the variable midpoint $\alpha_{123}\in U_{\alpha_1\alpha_2\alpha_3}$, even if we no longer indicate this explicitly.  Eventually one, two or perhaps all three of $\mathbb{S}_{\alpha_1}$, $\mathbb{S}_{\alpha_2}$ and $\mathbb{S}_{\alpha_3}$ may degenerate to a curve connecting the midpoint $\alpha_{123}$ with $\alpha_1$, $\alpha_2$ or $\alpha_3$, respectively. Whenever such is the case for all three surfaces, the closed trajectory $\mathbb{L}_{\alpha_1\alpha_2\alpha_3}$ cannot be expressed as the boundary of a 2--dimensional surface $\mathbb{S}_{\alpha_1\alpha_2\alpha_3}$. In what follows we will however exclude this latter possibility, so that at least one of the three surfaces on the right--hand side of (\ref{wdvdd}) does not degenerate to a curve.

In general we will not be able to compute the functional integral (\ref{xam}) exactly. However we can gain some insight from a steepest--descent approximation \cite{ZJ}, the details of which have been worked out in ref. \cite{NOI}. We find
\begin{equation}
g^{(0)}_{\alpha_1\alpha_2\alpha_3}(q_{\alpha_{123}})=\exp\left\{\frac{{\rm i}}{\hbar}S\left[\mathbb{L}^{(0)}_{\alpha_1\alpha_2\alpha_3}(\alpha_{123})\right]\right\},
\label{vamp}
\end{equation}
the superindex ${}^{(0)}$ standing for {\it evaluation at the extremal}. The latter is that path which, meeting the requirements stated after eqn. (\ref{fuga}),  minimises the action $S$. To summarise, by eqns. (\ref{xam}), (\ref{toes}), (\ref{wdvdd}), (\ref{bbvvdd}) and (\ref{vamp}), we can write the steepest--descent approximation to the 2--cocycle as
\begin{equation}
g_{\alpha_1\alpha_2\alpha_3}^{(0)}=\exp\left(\frac{{\rm i}}{\hbar}\int_{\mathbb{S}^{(0)}_{\alpha_1\alpha_2\alpha_3}}{\rm d}{\cal L}\wedge{\rm d}t\right),
\label{lecy}
\end{equation}
where $\mathbb{S}^{(0)}_{\alpha_1\alpha_2\alpha_3}$ is a minimal surface for the integrand ${\rm d}{\cal L}\wedge{\rm d}t$. We will henceforth drop the superindex ${}^{(0)}$, with the understanding that all our computations have been performed in the steepest--descent approximation.

\subsection{The connection}\label{nncc}

We can use eqns. (\ref{vamp}) and (\ref{lecy}) in order to compute the connection, at least to the same order of accuracy as the 2--cocycle itself. We find
\begin{equation}
A_{\alpha_1\alpha_2}=\frac{{\rm i}}{\hbar}\,\left({\cal L}\,{\rm d}t\right)_{\alpha_1\alpha_2},
\label{grag}
\end{equation}
\begin{equation}
B_{\alpha_2}-B_{\alpha_1}={\rm d}A_{\alpha_1\alpha_2}=\frac{{\rm i}}{\hbar}\,\left({\rm d}{\cal L}\wedge{\rm d}t\right)_{\alpha_1\alpha_2},
\label{mach}
\end{equation}
\begin{equation}
H\vert_{U_{\alpha}}={\rm d}B_{\alpha}.
\label{goed}
\end{equation}
A comment is in order. The potential $A$ is supposed to be a 1--form on configuration space $\mathbb{F}$, on which the gerbe is defined. As it stands in (\ref{grag}), due to the factor d$t$, $A$ is a 1--form on $\mathbb{F}\times\mathbb{R}$. If $\iota\colon\mathbb{F}\rightarrow\mathbb{F}\times\mathbb{R}$ denotes the natural inclusion, the 1--form $A$ in (\ref{grag}) is to be understood as its pullback $\iota^*({\cal L}{\rm d}t)$ onto $\mathbb{F}$. We will however continue to write it as ${\cal L}{\rm d}t$.

\section{Breaking  the U(1) symmetry on the gerbe}\label{bruch}

\subsection{Quantum mechanics as a gauge theory on a U(1) gerbe}\label{bernabeualberolachupamelapirola}

Let us perform the transformation
\begin{equation}
{\cal L}{\rm d}t\longrightarrow{\cal L}{\rm d}t+{\rm d}f,\qquad f\in C^{\infty}(\mathbb{F}),
\label{bbmj}
\end{equation}
where $f$ is an arbitrary function on $\mathbb{F}$ with the dimensions of an action. The above transformation does not alter the dynamics defined by $S$: it amounts to shifting $S$ by a constant $C$,
\begin{equation}
S\longrightarrow S+C,\qquad C:=\int{\rm d}f.
\label{chif}
\end{equation}
The way the transformation (\ref{bbmj}) acts on the quantum theory is well known. In the WKB approximation, the wavefunction reads \cite{GOSSON}
\begin{equation}
\psi_{\rm WKB}=R\exp\left(\frac{\rm i}{\hbar}S\right)
\label{ktfyrmlldmrd}
\end{equation}
for some amplitude $R$. Thus the transformation (\ref{bbmj}) multiplies the WKB wavefunction $\psi _{\rm WKB}$ and, more generally, any wavefunction $\psi$, by the constant phase factor $\exp\left({\rm i}{C}/{\hbar}\right)$:
\begin{equation}
\psi\longrightarrow \exp\left(\frac{{\rm i}}{{\hbar}}{C}\right)\psi.
\label{llkbkb}
\end{equation}
Gauging the rigid symmetry (\ref{llkbkb}) one obtains the transformation law 
\begin{equation}
\psi\longrightarrow\exp\left(\frac{{\rm i}}{{\hbar}}f\right)\psi, \qquad f\in C^{\infty}(\mathbb{F}),
\label{llmerk}
\end{equation}
$f$ being an arbitrary function on configuration space, with the dimensions of an action.  In the case of a gerbe over phase space,  as in ref. \cite{NOI}, the U(1) symmetry on the gerbe implies the possibility of performing the local gauge transformations (\ref{llmerk}). Analogous conclusions continue to hold in our case, where the gerbe is defined over configuration space $\mathbb{F}$; see  ref. \cite{NOI} for further details. In particular, for the transformation (\ref{llmerk}) to be an invariance of the theory, all derivatives within the action $S$ are to be covariantised by means of the connection 1--form $A$ of eqn. (\ref{grag}).

\subsection{A gerbe analogue of the Higgs mechanism}\label{hicks}

The U(1) symmetry (\ref{llmerk}) on the gerbe is spontaneously broken. If this symmetry were unbroken, then in particular the duality (\ref{bbmk}) between the semiclassical and the strong quantum regimes, or its equivalent (\ref{paddy}) between long and short distances, would be manifest. This is certainly not the case in Nature as observed at low energies, although it has been suggested \cite{PADUNO} that effects such as the dualities (\ref{paddy}) and (\ref{bbmk}) are to be expected within quantum gravity. The breaking occurs via a mechanism that is analogous to the Higgs mechanism of Yang--Mills theory. However, since gerbes fall into a category that is {\it one step up}\/ from that of fibre bundles, the details of the symmetry--breaking mechanism are different here. For the breaking to take place, a certain field must develop a vacuum expectation value equal to Planck's constant. This is so because quantisation is due to a {\it nonvanishing}\/ value for $\hbar$, and we are interpreting quantum mechanics as a gauge theory. Moreover, whatever nonvanishing value $\hbar$ may take on, different numerical values for Planck's constant lead to different quantum theories. A specific choice of one particular value for $\hbar$ picks one, and only one, quantum theory out of the many that are possible before the U(1) symmetry is broken.

Consider the connection 1--form $\hat A$ on the gerbe. As usual, the caret reminds us  that $\hat A$ is a quantum operator corresponding to the classical field $A$. By eqn. (\ref{grag}), $\hat A$ is proportional to the operator $\hat{\cal L}{\rm d}t$. The expectation value $\langle\hat{\cal L}{\rm d}t\rangle$ can be obtained as the integral over a certain path $\mathbb{L}_i$, the latter playing the role of a certain vacuum state:
\begin{equation}
\langle\hat{\cal L}{\rm d}t\rangle_{\mathbb{L}_i}:=\int_{\mathbb{L}_i}{\cal L}{\rm d}t=\hbar_i.
\label{phone}
\end{equation}
In principle each path $\mathbb{L}_i$, or vacuum state, produces a different value for $\hbar_i$.  Because Feyman's kernel is ${\rm exp}\left({\rm i}S/\hbar\right)$,
before symmetry breaking there is a whole U(1)'s worth of equivalent vacua. Now the vacuum state $\mathbb{L}_{\rm phys}$ actually picked by Nature gives rise to the physical value $\hbar_{\rm phys}$ of Planck's constant as observed in our world:
\begin{equation}
\langle\hat{\cal L}{\rm d}t\rangle_{\mathbb{L}_{\rm phys}}=\int_{\mathbb{L}_{\rm phys}}{\cal L}{\rm d}t=\hbar_{\rm phys}.
\label{phonz}
\end{equation}
The corresponding $\mathbb{L}_{\rm phys}$ must have a length $\sim O(L_P)$. Our notations stress the difference between $\hbar_i$ as a variable parameter and $\hbar_{\rm phys}$, the latter being the particular value assumed by that parameter in the actual world we live in. Eqn. (\ref{phonz}) expresses the breaking of the U(1) symmetry on the gerbe.

We can also recast (\ref{phone}) and (\ref{phonz}) in terms of surfaces $\mathbb{S}$ and 2--forms:
\begin{equation}
\langle{\rm d}\hat{\cal L}\wedge{\rm d}t\rangle_{\mathbb{S}_i}:=\int_{\mathbb{S}_i}{\rm d}{\cal L}\wedge{\rm d}t=\hbar_i.
\label{higgins}
\end{equation}
Again each surface $\mathbb{S}_i$, or vacuum state, produces a different value for $\hbar_i$. Also, the vacuum state $\mathbb{S}_{\rm phys}$ actually picked by Nature must have an area $\sim O(L_P^2)$ and be such that
\begin{equation}
\langle{\rm d}\hat{\cal L}\wedge{\rm d}t\rangle_{\mathbb{S}_{\rm phys}}=\int_{\mathbb{S}_{\rm phys}}{\rm d}{\cal L}\wedge{\rm d}t=\hbar_{\rm phys}.
\label{higginz}
\end{equation}
By eqn. (\ref{mach}), the above can also be expressed in terms of the Neveu--Schwarz operator 2--form $\hat B$. If the surfaces $\mathbb{S}_i$ have boundaries $\partial\mathbb{S}_i=\mathbb{L}_i$, then eqns. (\ref{higginz}) and (\ref{higgins}) are strictly equivalent to (\ref{phonz}) and (\ref{phone}), respectively. However the convenience of using surfaces $\mathbb{S}$ rather than paths $\mathbb{L}$ will become clear presently. We conclude that {\it a nonvanishing value for the quantum of action $\hbar$ is equivalent to a nonvanishing quantum of length proportional to $L_P$, or to a nonvanishing quantum of area proportional to $L_P^2$}.  

We started off with a configuration space $\mathbb{F}$ whose coordinates $q^j$ were commutative. Next we constructed a U(1) gerbe over $\mathbb{F}$. The U(1) symmetry on the latter allowed us to arbitrarily pick, on a point--by--point basis, the zero point for the action integral $S$. As proved in ref. \cite{NOI} and summarised in section \ref{bernabeualberolachupamelapirola}, this symmetry rendered notions like {\it semiclassical approximation}\/ or {\it strong--quantum regime}\/ meaningless. Finally we observed that the U(1) symmetry  must be spontaneously broken at low energies, where the above notions {\it do}\/ have a definite meaning. A quantum of area results as a consequence, which can only exist on a noncommutative space. We can therefore state that {\it the Higgs mechanism on the U(1) gerbe over configuration space $\mathbb{F}$ renders the latter noncommutative}.

On the other hand, as observed at low energies, space coordinates are definitely commutative, while they are expected to turn noncommutative at an energy scale around that of quantum gravity. The whole situation can be summarised in the diagram
\begin{equation}
\mathbb{F}( \hbar)\longrightarrow\mathbb{F}_{\star}(\hbar=\hbar_{\rm phys})\longrightarrow\mathbb{F}(\hbar_{\rm phys}\to 0).
\label{vite}
\end{equation}
The first arrow stands for the Higgs mechanism described above. It represents the passage from the commutative configuration space $\mathbb{F}(\hbar)$, where no value for $\hbar$ has been specified yet, to the noncommutative space $\mathbb{F}_{\star}(\hbar=\hbar_{\rm phys})$, on which a specific value $\hbar_{\rm phys}$ for $\hbar$ has been selected.  The ${\star}$ in the notation  stresses the fact that the multiplication law now is the noncommutative $\star$--product \cite{NCG}. The second arrow represents the passage to the limit $\hbar_{\rm phys}\to 0$, in which the $\star$--product on  $\mathbb{F}_{\star}$ becomes the usual, pointwise, commutative multiplication law on the commutative space $\mathbb{F}$. This is the passage from the high--energy world $\mathbb{F}_{\star}(\hbar=\hbar_{\rm phys})$, where quantum--gravity effects are expected to be relevant, to the low--energy world $\mathbb{F}(\hbar_{\rm phys}\to 0)$ we live in, where such effects can be neglected.

\subsection{The uncertainty principle on configuration space}\label{unbestt}

Strictly speaking, a quantum of area makes sense only on a noncommutative space; commutative continua do not allow for such a coarse graining, since infinitesimals can be made as small as one pleases. Therefore {\it a nonvanishing quantum of area is a consequence of the nonvanishing of the noncommutativity parameter $\theta^{ij}$}. The uncertainty principle (\ref{rkbr}) on configuration space then follows immediately. This is where one advantage of using surfaces rather than loops becomes apparent: by eqn. (\ref{mach}), the vacuum expectation value (\ref{higginz}) can be related to the vacuum expectation value of the Neveu--Schwarz 2--form operator $\hat B$. A choice of gauge ensures $\hat B_{\alpha_1}=0$, while the caret can be removed by integrating over the surface $\mathbb{S}_{\rm phys}$.  If we take the latter as spanning the $j$, $k$ spatial dimensions, this integral is proportional to $\theta^{-1}_{jk}$, which is the (inverse) noncommutativity parameter.  

However, there is one fundamental difference between the uncertainty principle on configuration space and the usual uncertainty principle on phase space. Namely, the latter is the result of rewriting the classical Poisson brackets $\{q,p\}=1$ in terms of quantum commutators, while the uncertainty principle on configuration space is a consequence of the breaking of the U(1) symmetry on the gerbe. Thus, while Heisenberg's principle on phase space follows from the {\it kinematic}\/ equation $[\hat q,\hat p]={\rm i}\hbar$, the uncertainty principle (\ref{rkbr}) on configuration space involves $\hbar_{\rm phys}$ as a {\it dynamically generated quantum scale}. As such, Planck's constant 
will be subject to a renormalisation--group law, like the Yang--Mills coupling constant. In particular, the value of $\hbar$ may depend on the scale. This is in perfect agreement with the conclusions of ref.  \cite{FARAGGIMATONE} regarding Planck's constant, and also with those of ref. \cite{REUTER} regarding Newton's constant.

To finish this section we would like to comment on the Higgs mechanism on phase space. In the limit $L_P\to 0$, configuration space becomes commutative, while phase space retains a nonvanishing commutator (or Poisson brackets) between coordinates and momenta. This is so because $[\hat q^j, \hat q^k]\sim L_P^2$, while $[\hat q^j, \hat p^k]$ is order zero in $L_P$.

\subsection{The characteristic class}\label{klasse}

Next we would like to relate area quantisation to the quantised characteristic class for the gerbe.

It follows from eqn. (\ref{mach}) that ${\rm d}B_{\alpha_1}={\rm d}B_{\alpha_2}$. This implies that the 3--form field strength $H$, contrary to the 2--form potential $B$, is globally defined on $\mathbb{F}$. Now the de Rham cohomology class $[H]$ of the 3--form $H$ is quantised \cite{ORLANDO}:
\begin{equation}
[H]\in H^3(\mathbb{F}, 2\pi{\rm i}\mathbb{Z}),
\label{cuent}
\end{equation}
{\it i.e.},
\begin{equation}
\frac{1}{2\pi{\rm i}}\int_{\mathbb{V}'}H\in\mathbb{Z}, \qquad \partial\mathbb{V'}=0,
\label{eqqwe}
\end{equation}
for all 3--dimensional volumes $\mathbb{V'}\subset\mathbb{F}$ such that $\partial\mathbb{V'}=0$.

Consider now a 3--dimensional volume $\mathbb{V}\subset\mathbb{F}$ whose boundary is a 2--dimensional closed surface $\mathbb{S}$. If $\mathbb{V}$ is connected and simply connected we may, without loss of generality, take $\mathbb{V}$ to be a solid ball, so $\mathbb{S}=\partial\mathbb{V}$ is a sphere. Let us cover $\mathbb{S}$ by stereographic projection.  This gives us two coordinate charts, respectively centred around the north and south poles on the sphere. Each chart is diffeomorphic to a copy of the plane $\mathbb{R}^2$. Each plane covers the whole sphere $\mathbb{S}$ with the exception of the opposite pole. The intersection of these two charts is the whole sphere $\mathbb{S}$ punctured at its north and south poles. Let us embed the chart $\mathbb{R}^2_{\alpha_1}$ centred at the north pole within the open set $ U_{\alpha_1}$, {\it i.e.}, $\mathbb{R}^2_{\alpha_1}\subset U_{\alpha_1}$, if necessary by means of some diffeomorphism. Analogously, for the south pole we have $\mathbb{R}^2_{\alpha_2}\subset U_{\alpha_2}$. There is also no loss of generality in assuming that only two points on the sphere $\mathbb{S}$ (the north and south poles) remain outside the 2--fold overlap $U_{\alpha_1}\cap U_{\alpha_2}$. By Stokes' theorem,
\begin{equation}
\int_{\mathbb{V}}H=\int_{\mathbb{V}}{\rm d}B=\int_{\partial\mathbb{V}}B=\int_{\mathbb{S}}B=\int_{\mathbb{R}^2_{\alpha_2}}B-\int_{\mathbb{R}^2_{\alpha_1}}B,
\label{kote}
\end{equation}
and, by eqn. (\ref{mach}),
\begin{equation}
\int_{\mathbb{V}}H=\frac{{\rm i}}{\hbar}\int_{\mathbb{R}^2-\{0\}}{\rm d}{\cal L}\wedge{\rm d}t,
\label{kabaka}
\end{equation}
where $\mathbb{R}^2-\{0\}$ denotes either one of our two charts, punctured at its corresponding origin. Now $\mathbb{R}^2-\{0\}$ falls short of covering the whole sphere $\mathbb{S}$ by just two points (the north and south poles), and the latter have zero measure. Excluding cases where the integrand is supported on isolated points such as the poles, we may just as well write
\begin{equation}
\int_{\mathbb{V}}H=\frac{{\rm i}}{\hbar}\int_{\mathbb{S}}{\rm d}{\cal L}\wedge{\rm d}t, \qquad \partial\mathbb{V}=\mathbb{S}.
\label{kakaka}
\end{equation}
Eqn. (\ref{kakaka}) is analogous to the Gauss law in electrostatics, with $H$ replacing the electric charge density 3--form and ${\rm i}{\rm d}{\cal L}\wedge{\rm d}t/\hbar$ replacing the corresponding surface flux 2--form. 

Now the quantisation condition (\ref{cuent}) on $[H]$ applies to {\it closed}\/ volumes, while (\ref{kakaka}) refers to volumes bounded by a surface. However it seems reasonable to conjecture that (\ref{cuent}) should be related to some quantisation condition on the surface integral of ${\rm i}{\rm d}{\cal L}\wedge{\rm d}t/\hbar$. Since the surface integral of ${\rm i}{\rm d}{\cal L}\wedge{\rm d}t/\hbar$ is related to that of the Neveu--Schwarz field $B$, the vacuum expectation value for $\hat B$ will be quantised: a fact that is already known to us from the foregoing discussion.  Specifically: if one postulates the quantisation condition
\begin{equation}
\frac{{\rm i}}{\hbar}[{\rm d}{\cal L}\wedge{\rm d}t]\in H^2(\mathbb{F},\pi{\rm i}\mathbb{Z}),
\label{cojo}
\end{equation}
then the quantisation condition (\ref{cuent}) follows, and viceversa. To prove this, consider two volumes $\mathbb{V}_1$,  $\mathbb{V}_2$ such that $\partial\mathbb{V}_1=\mathbb{S}=-\partial\mathbb{V}_2$, and such that glued together along their common boundary one obtains a $\mathbb{V}'$ without boundary. Then
$$
\frac{1}{\pi\hbar}\int_{\mathbb{S}}{\rm d}{\cal L}\wedge{\rm d}t=\frac{1}{2\pi\hbar}\int_{\mathbb{S}}{\rm d}{\cal L}\wedge{\rm d}t+\frac{1}{2\pi\hbar}\int_{\mathbb{S}}{\rm d}{\cal L}\wedge{\rm d}t
$$
$$
=\frac{1}{2\pi\hbar}\int_{\partial\mathbb{V}_1}{\rm d}{\cal L}\wedge{\rm d}t-\frac{1}{2\pi\hbar}\int_{\partial\mathbb{V}_2}{\rm d}{\cal L}\wedge{\rm d}t
$$
\begin{equation}
=\frac{1}{2\pi{\rm i}}\int_{\mathbb{V}_1}H+\frac{1}{2\pi{\rm i}}\int_{\mathbb{V}_2}H=\frac{1}{2\pi{\rm i}}\int_{\mathbb{V}'}H.
\label{kath}
\end{equation}
Now the last term above is an integer if and only if also the first term is an integer. This proves that (\ref{cojo}) and (\ref{cuent}) are equivalent: area quantisation on configuration space and a quantised characterictic class for the gerbe are equivalent statements.

We can return to eqn. (\ref{lecy}) and rewrite the 2--cocycle using (\ref{cojo}):
\begin{equation}
g_{\alpha_1\alpha_2\alpha_3}=\exp\left({\rm i}\pi n_{\alpha_1\alpha_2\alpha_3}\right),\qquad n_{\alpha_1\alpha_2\alpha_3}\in\mathbb{Z}.
\label{lecyxx}
\end{equation}
In the particular case of the free particle we conclude, by (\ref{bonax}) below, that $n_{\alpha_1\alpha_2\alpha_3}$ must be even: $n_{\alpha_1\alpha_2\alpha_3}=2k_{\alpha_1\alpha_2\alpha_3}$, for some $k_{\alpha_1\alpha_2\alpha_3}\in\mathbb{Z}$.

\appendix

\section{Appendix: computing the trivialisation}\label{bernabeumarikondemierda}

Let us work out the trivialisation explicitly for the case of a particle on the manifold $\mathbb{F}\times\mathbb{R}$. Assume that local charts on $\mathbb{F}$ are diffeomorphic to $\mathbb{R}^{d}$, where $d=n-1$. This simplifying assumption allows one to perform all computations explicitly. In what follows, our propagators are normalised as in ref. \cite{DR}. However it must be borne in mind that we are interested only in the nonconstant U(1)--valued phase of the final result.

\subsection{The constant potential}\label{kkonn}

The propagator for a free particle is
\begin{equation}
G_0(q_{\alpha_1}, t_{\alpha_1};q_{\alpha_2}, t_{\alpha_2})=\left[\frac{m}{2\pi{\rm i}\hbar(t_{\alpha_2}-t_{\alpha_1})}\right]^{d/2}\exp\left[\frac{{\rm i}m}{2\hbar}\frac{(q_{\alpha_2}-q_{\alpha_1})^2}{t_{\alpha_2}-t_{\alpha_1}}\right].
\label{prop}
\end{equation}
By eqn. (\ref{bona}) 
$$
\tilde\tau_{\alpha_1\alpha_2}=\int{\rm d}q_{\alpha_1}{\rm d}q_{\alpha_2}G_0(q_{\alpha_1}, t_{\alpha_1};q_{\alpha_{12}}, t_{\alpha_{12}})G_0(q_{\alpha_{12}}, t_{\alpha_{12}};q_{\alpha_2}, t_{\alpha_2})
$$
$$
=\left[\frac{m}{2\pi{\rm i}\hbar(t_{\alpha_{12}}-t_{\alpha_1})}\right]^{d/2}\int{\rm d}q_{\alpha_1}\exp\left[\frac{{\rm i}m}{2\hbar}\frac{(q_{\alpha_{12}}-q_{\alpha_1})^2}{t_{\alpha_{12}}-t_{\alpha_1}}\right]
$$
\begin{equation}
\times\left[\frac{m}{2\pi{\rm i}\hbar(t_{\alpha_2}-t_{\alpha_{12}})}\right]^{d/2}\int{\rm d}q_{\alpha_2}\exp\left[\frac{{\rm i}m}{2\hbar}\frac{(q_{\alpha_2}-q_{\alpha_{12}})^2}{t_{\alpha_2}-t_{\alpha_{12}}}\right]
=1.
\label{bonax}
\end{equation}
Hence the free particle has a trivial, {\it i.e.}, constant, trivialisation.

\subsection{The linear potential}\label{llinn}

Consider a particle acted on by a constant force $F$. The propagator then reads 
\begin{equation}
G_1(q_{\alpha_1}, t_{\alpha_1};q_{\alpha_2}, t_{\alpha_2})=\left[\frac{m}{2\pi{\rm i}\hbar(t_{\alpha_2}-t_{\alpha_1})}\right]^{d/2}
\label{propxx}
\end{equation}
$$
\times\exp\left\{\frac{{\rm i}}{\hbar}\left[\frac{m}{2}\frac{(q_{\alpha_2}-q_{\alpha_1})^2}{t_{\alpha_2}-t_{\alpha_1}}+\frac{F}{2}(t_{\alpha_2}-t_{\alpha_1})(q_{\alpha_2}+q_{\alpha_1})-\frac{F^2}{24m}(t_{\alpha_2}-t_{\alpha_1})^3\right]\right\}.
$$
By eqn. (\ref{bona}) 
$$
\tilde\tau_{\alpha_1\alpha_2}=\int{\rm d}q_{\alpha_1}{\rm d}q_{\alpha_2}G_1(q_{\alpha_1}, t_{\alpha_1};q_{\alpha_{12}}, t_{\alpha_{12}})G_1(q_{\alpha_{12}}, t_{\alpha_{12}};q_{\alpha_2}, t_{\alpha_2})
$$
\begin{equation}
=\left[\frac{m}{2\pi{\rm i}\hbar(t_{\alpha_{12}}-t_{\alpha_1})}\right]^{d/2}\left[\frac{m}{2\pi{\rm i}\hbar(t_{\alpha_2}-t_{\alpha_{12}})}\right]^{d/2}J_{\alpha_1}J_{\alpha_2},
\label{bernabeucalabaceu}
\end{equation}
where the integrals $J_{\alpha_1}$ and $J_{\alpha_2}$ are defined as
\begin{equation}
J_{\alpha_1}:=
\int{\rm d}q_{\alpha_1}\exp\left[\frac{{\rm i} m}{2\hbar}\frac{(q_{\alpha_{12}}-q_{\alpha_1})^2}{t_{\alpha_{12}}-t_{\alpha_{1}}}\right]
\label{bernabeuketehostien}
\end{equation}
$$
\times\exp\left\{\frac{{\rm i}}{\hbar}\left[\frac{F}{2}(t_{\alpha_{12}}-t_{\alpha_{1}})(q_{\alpha_{12}}+q_{\alpha_{1}})-\frac{F^2}{24m}(t_{\alpha_{12}}-t_{\alpha_{1}})^3\right]\right\}
$$
and
\begin{equation}
J_{\alpha_2}:=\int{\rm d}q_{\alpha_2}\exp\left[\frac{{\rm i} m}{2\hbar}\frac{(q_{\alpha_2}-q_{\alpha_{12}})^2}{t_{\alpha_2}-t_{\alpha_{12}}}\right]
\label{bernabeuketefollen}
\end{equation}
$$
\times\exp\left\{\frac{{\rm i}}{\hbar}\left[\frac{F}{2}(t_{\alpha_{2}}-t_{\alpha_{12}})(q_{\alpha_2}+q_{\alpha_{12}})-\frac{F^2}{24m}(t_{\alpha_2}-t_{\alpha_{12}})^3\right]\right\}.
$$
Now the integrals (\ref{bernabeuketehostien}) and (\ref{bernabeuketefollen}) are readily evaluated, with the results
\begin{equation}
J_{\alpha_1}=\left[\frac{2\pi{\rm i}\hbar(t_{\alpha_{12}}-t_{\alpha_1})}{m}\right]^{d/2}\exp\left\{\frac{{\rm i}}{\hbar}\left[-\frac{F^2}{6m}(t_{\alpha_{12}}-t_{\alpha_1})^3+Fq_{\alpha_{12}}(t_{\alpha_{12}}-t_{\alpha_1})\right]\right\}
\label{bernabeumarikonzon}
\end{equation}
and
\begin{equation}
J_{\alpha_2}=\left[\frac{2\pi{\rm i}\hbar(t_{\alpha_2}-t_{\alpha_{12}})}{m}\right]^{d/2}
\exp\left\{\frac{{\rm i}}{\hbar}\left[-\frac{F^2}{6m}(t_{\alpha_2}-t_{\alpha_{12}})^3+Fq_{\alpha_{12}}(t_{\alpha_2}-t_{\alpha_{12}})\right]\right\}.
\label{jotados}
\end{equation}
{}Finally substituting the integrals  (\ref{bernabeumarikonzon}) and (\ref{jotados}) into eqn. (\ref{bernabeucalabaceu}) we obtain the trivialisation
$$
\tau_{\alpha_1\alpha_2}=\exp\left\{\frac{{\rm i}}{\hbar}\left[-\frac{F^2}{6m}(t_{\alpha_{12}}-t_{\alpha_1})^3+Fq_{\alpha_{12}}(t_{\alpha_{12}}-t_{\alpha_1})\right]\right\}
$$
\begin{equation}
\times\exp\left\{\frac{{\rm i}}{\hbar}\left[-\frac{F^2}{6m}(t_{\alpha_2}-t_{\alpha_{12}})^3+Fq_{\alpha_{12}}(t_{\alpha_2}-t_{\alpha_{12}})\right]\right\}.
\label{mierd}
\end{equation}
Eqn.  (\ref{mierd}) correctly reduces to the free--particle trivialisation (\ref{bonax}) when $F=0$.

\subsection{The quadratic potential}\label{qqddr}

As a final example we will work out the trivialisation for an isotropic harmonic oscillator with frequency $\omega$. Here the propagator is given by
$$
G_2(q_{\alpha_1}, t_{\alpha_1};q_{\alpha_2}, t_{\alpha_2})=\left\{\frac{m\omega}{2\pi{\rm i}\hbar\sin[\omega(t_{\alpha_2}-t_{\alpha_1})]}\right\}^{d/2}
$$
\begin{equation}
\times\exp\left(\frac{{\rm i}m\omega}{2\hbar\sin[\omega(t_{\alpha_2}-t_{\alpha_1})]}\left\{(q_{\alpha_1}^2+q_{\alpha_2}^2)\cos[\omega(t_{\alpha_2}-t_{\alpha_1})]-2q_{\alpha_1}q_{\alpha_2}\right\}\right).
\label{osk}
\end{equation}
Again by eqn. (\ref{bona}) 
$$
\tilde\tau_{\alpha_1\alpha_2}=\int{\rm d}q_{\alpha_1}{\rm d}q_{\alpha_2}G_2(q_{\alpha_1}, t_{\alpha_1};q_{\alpha_{12}}, t_{\alpha_{12}})G_2(q_{\alpha_{12}}, t_{\alpha_{12}};q_{\alpha_2}, t_{\alpha_2})
$$
\begin{equation}
=\left\{\frac{m\omega}{2\pi{\rm i}\hbar\sin[\omega(t_{\alpha_{12}}-t_{\alpha_1})]}\right\}^{d/2}
\left\{\frac{m\omega}{2\pi{\rm i}\hbar\sin[\omega(t_{\alpha_2}-t_{\alpha_{12}})]}\right\}^{d/2}
K_{\alpha_1}K_{\alpha_2},
\label{bernabeukaka}
\end{equation}
where the integrals $K_{\alpha_1}$ and $K_{\alpha_2}$ are defined by
\begin{equation}
K_{\alpha_1}:=
\label{bernabeubastardeu}
\end{equation}
$$
\int{\rm d}q_{\alpha_1}\exp\left(\frac{{\rm i}m\omega}{2\hbar\sin[\omega(t_{\alpha_{12}}-t_{\alpha_1})]}\left\{(q_{\alpha_1}^2+q_{\alpha_{12}}^2)\cos[\omega(t_{\alpha_{12}}-t_{\alpha_1})]-2q_{\alpha_1}q_{\alpha_{12}}\right\}\right)
$$
and
\begin{equation}
K_{\alpha_2}:=
\label{bernabeubastardeus}
\end{equation}
$$
\int{\rm d}q_{\alpha_2}\exp\left(\frac{{\rm i}m\omega}{2\hbar\sin[\omega(t_{\alpha_{2}}-t_{\alpha_{12}})]}\left\{(q_{\alpha_{12}}^2+q_{\alpha_{2}}^2)\cos[\omega(t_{\alpha_{2}}-t_{\alpha_{12}})]-2q_{\alpha_{12}}q_{\alpha_{2}}\right\}\right).
$$
One finds
\begin{equation}
K_{\alpha_1}=\left\{\frac{2\pi {\rm i}\hbar}{m\omega}\tan\left[\omega(t_{\alpha_{12}}-t_{\alpha_1})\right]\right\}^{d/2}\exp\left\{-\frac{{\rm i}m\omega}{2\hbar}\tan\left[\omega(t_{\alpha_{12}}-t_{\alpha_1})\right]q_{\alpha_{12}}^2\right\}
\label{kagonpepebernabeu}
\end{equation}
and
\begin{equation}
K_{\alpha_2}=\left\{\frac{2\pi {\rm i}\hbar}{m\omega}\tan\left[\omega(t_{\alpha_{2}}-t_{\alpha_{12}})\right]\right\}^{d/2}\exp\left\{-\frac{{\rm i}m\omega}{2\hbar}\tan\left[\omega(t_{\alpha_{2}}-t_{\alpha_{12}})\right]q_{\alpha_{12}}^2\right\}.
\label{kagonpepeubernabeu}
\end{equation}
Substituting eqns. (\ref{kagonpepebernabeu}) and (\ref{kagonpepeubernabeu}) into eqn. (\ref{bernabeukaka}), normalising by the corresponding modulus and dropping all constant phase factors we obtain the trivialisation
\begin{equation}
\tau_{\alpha_1\alpha_2}=\exp\left(-\frac{{\rm i}m\omega}{2\hbar}\left\{\tan\left[\omega(t_{\alpha_{12}}-t_{\alpha_1})\right]+ \tan\left[\omega(t_{\alpha_{2}}-t_{\alpha_{12}})\right]\right\}q_{\alpha_{12}}^2\right).
\label{bernabeuketedenporelaneu}
\end{equation}
Eqn.  (\ref{bernabeuketedenporelaneu}) also reduces to the free--particle trivialisation (\ref{bonax}) when $\omega=0$.

\break
\hfill

\centerline{\bf Figure}

\centerline{The closed trajectory $\mathbb{L}_{\alpha_1\alpha_2\alpha_3}$ of eqn. (\ref{fuga}).}

\includegraphics[scale=0.6, bb= -2cm 16cm -2cm 0cm]{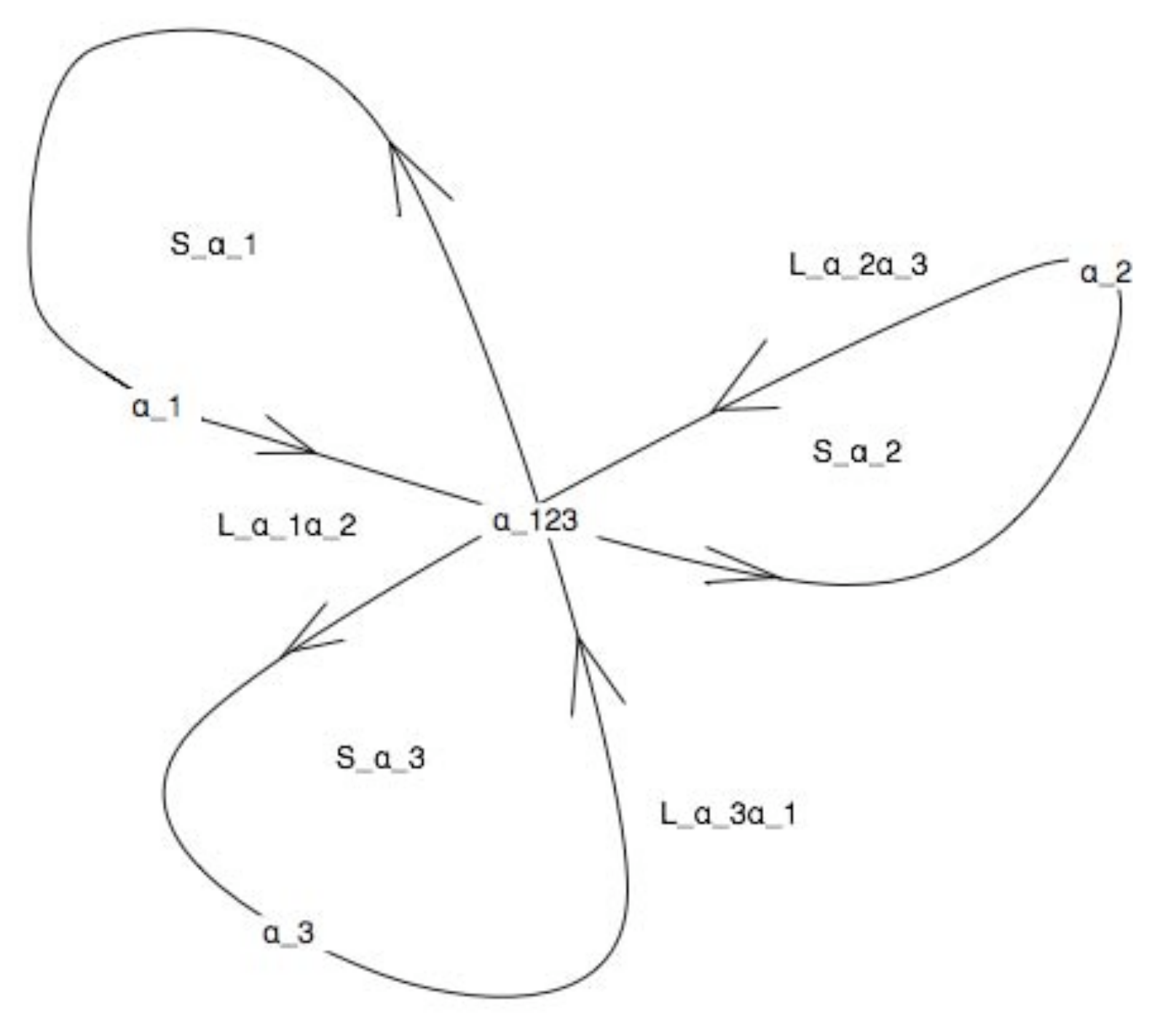}
\vskip12cm

\noindent {\bf Acknowledgements}  It is a great pleasure to thank Max--Planck--Institut f\"ur Gravitationsphysik, Albert--Einstein--Institut (Golm, Germany) for hospitality during the preparation of this article. This work has been supported by Ministerio de Educaci\'on y Ciencia through grant FIS2005--02761 and by Generalitat Valenciana (Spain).

\break

\end{document}